\documentclass[11pt,twoside,dvips]{article}

\usepackage{graphicx}
\usepackage{asp2006}
\usepackage{epsf}
\usepackage{psfig}
\usepackage{lscape}

\begin{document}

\def\MSUN{\rm M_{\odot}}
\def\RSUN{\rm R_{\odot}}
\def\MSUNYR{\rm M_{\odot}\,yr^{-1}}
\def\MDOT{\dot{M}}

\newbox\grsign \setbox\grsign=\hbox{$>$} \newdimen\grdimen \grdimen=\ht\grsign
\newbox\simlessbox \newbox\simgreatbox
\setbox\simgreatbox=\hbox{\raise.5ex\hbox{$>$}\llap
     {\lower.5ex\hbox{$\sim$}}}\ht1=\grdimen\dp1=0pt
\setbox\simlessbox=\hbox{\raise.5ex\hbox{$<$}\llap
     {\lower.5ex\hbox{$\sim$}}}\ht2=\grdimen\dp2=0pt
\def\simgreat{\mathrel{\copy\simgreatbox}}
\def\simless{\mathrel{\copy\simlessbox}}

\title{Outflows from AGN: their Impact on Spectra and the Environment}   
%%% Fill in title
\author{Daniel Proga and Ryuichi Kurosawa$^*$}   %%% Fill in author names
\affil{Department of Physics and Astronomy, University of Nevada Las Vegas, NV 89154, USA\\
 $^*$Current Address: Department of Astronomy, Cornell University, Ithaca, NY 1\
4853, USA} % additional visiting address}                                                

\begin{abstract} %%% Abstract to run on from here.
We present a brief summary of the main results from our multi-dimensional, 
time-dependent simulations of gas dynamics in AGN. We focus on two types of 
outflows powered by radiation emitted from the AGN:
disk winds and winds driven from large-scale inflows.
We show spectra predicted by the simulations and discuss their relevance
to observations of broad- and narrow-line regions of the AGN. We finish 
with a few remarks on whether these outflows can have a significant impact 
on their environment and host galaxy.

\end{abstract}

%%% MAIN BODY OF TEXT GOES HERE. CONSULT "INSTRUCTIONS FOR AUTHORS USING
%%% LATEX2E MARKUP", SECTIONS 2.3-2.6 FOR HELP WITH EQUATIONS, FIGURES,
%%% AND TABLES.

%\section{}   %%% Top level section head (remove "%" symbol)
%\subsection{}   %%% Second level section head (remove "%" symbol)
%\subsubsection{}   %%% Lowest level section head (remove "%" symbol)
%\section*{}    %%% Unnumbered top level section head (remove "%" symbol)
%\subsection*{}   %%% Unnumbered second level section head (remove "%" symbol)

\section{Introduction}

While AGN are very strong sources of electromagnetic radiation,
they are also sources of
outflows of matter and magnetic energy. There is a growing body of evidence
that outflows are quite common and are an integral part of AGN
\citep[e.g.,][]{Crenshaw:2002, Richards:2004}.  Moreover, radiation and
mass outflows in luminous sources are physically coupled as
matter can absorber, scatter and emit radiation and
radiation can affect dynamics and ionization of
matter. For example, powerful mass outflows in quasars are
very likely winds driven by radiation
from accretion disks  \citep[e.g., reviews by][]{Konigl:2006, Proga:2007a}.
In addition, as we will illustrate here, the same radiation
can drive outflows from large scale inflows.

In this paper, we briefly summarize results from numerical simulations
of disk winds and large-scale outflows. We review the properties
of these outflows and compare them with those observed in AGN.
This will allow us to assess the level of our understanding of
AGN outflows and their potential role in the AGN feedback.

\section{Radiation-Driven Outflows from Accretion Disks}

The successes of modelling outflows driven by radiation pressure
on spectra lines (line driving) from
OB stars \citep[e.g.,][]{Castor:1975, Puls:2008} 
and from accretion disks in cataclysmic variables 
\citep[CVs; e.g.,][]{Proga:2005}
motivate applications of a similar physics to model outflows
in AGN.

\begin{figure}[!ht]
  \begin{center}
    \includegraphics[angle=90, width=0.6\textwidth]{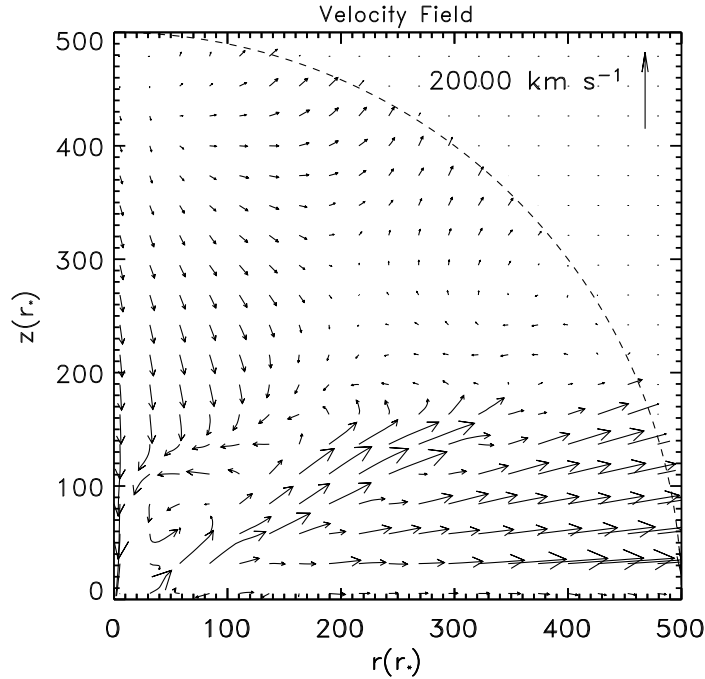}
  \end{center}
\caption{
{\small 
A map of the poloidal velocity field 
of the radiation-driven disk wind model described in the text. 
The rotation axis of the disk is along the left hand vertical
frame, while the midplane of the disk is along the lower horizontal frame.
The figure is from \citet{Proga:2004}.
}}
\end{figure}

In \citet{Proga:2000} and \citet{Proga:2004},  we adopted the approach
from \citet{Proga:1998,Proga:1999a} 
to calculate  axisymmetric time-dependent hydrodynamical
models of line-driven  winds from accretion 
disks in AGN. To apply the line-driven  disk wind model developed for CVs to AGN, we
took into account, in an approximated way, 
the difference in the spectral energy distribution 
between AGN and CVs. In particular, we introduced
three major modifications to our previous approach:
1) calculation of the parameters of the line force
based on the wind 
properties, 2) effects of optical depth on the continuum photons,
and 3) radiative heating and cooling of the gas.

Our AGN wind calculations follow 
(i) a hot and low density flow with negative radial 
velocity in the polar region, (ii) a dense, warm and fast {\it equatorial}  
outflow from the disk, and (iii) a transitional zone in which the disk outflow 
is hot and struggles to escape the system. 
Fig.~1 shows an example of a velocity field of a AGN disk wind model.

To make a direct comparison with observations, 
we calculated synthetic line profiles based on our model for
the C~IV $\lambda$1549~\AA~line \citep{Proga:2004}.
The synthetic line profiles show a strong dependence on inclination 
angle: the absorption forms only when 
an observer looks at the CE through the fast wind (i.e., 
$i \geq 60^\circ$ as shown in Fig.~3 of \citealt{Proga:2004};
see also Fig.~2 here). This $i$
dependence can explain why only 10\% of QSO have BALs. 

\begin{figure}[!ht]
\begin{center}
\includegraphics[width=13cm,angle=0]{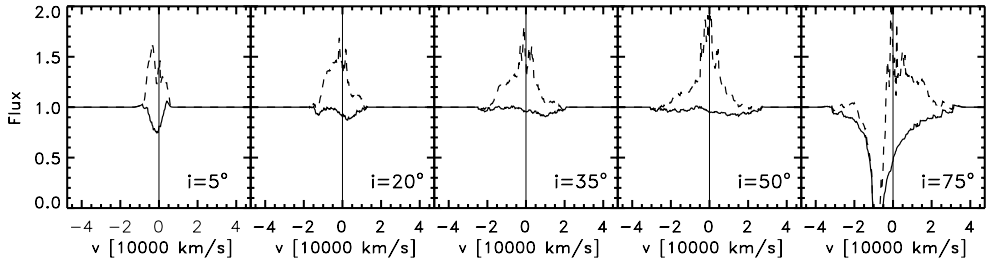}
\caption{
{\small 
Theoretical profiles  of C{\sc iv}~1549\AA\ based on
the wind model of  \citet{Proga:2004}  as a function of inclination angle,
$i$ (see bottom right corner of each panel for the value of $i$).
The solid lines show the profiles due to the absorption only
whereas the dashed lines show the profiles due to the absorption
and emission (the line source function includes resonance scattering
and thermal emission).
Note the sensitivity of the lines on the inclination angle.
The zero velocity corresponding to the line center is indicated
by the vertical line.
}}
\end{center}
\end{figure}

The model also predicts high column densities for the 
inclination angles at which strong absorption lines form. 
\citet{Schurch:2009} computed broad band spectra for various inclination angles
using the simulation from \citet{Proga:2004}. Fig.~3 presents some
examples and shows that the model is 
consistent with the observations, i.e., BAL QSO are
underluminous in X-rays compared to their non-BAL QSO counterparts
\citep[e.g.,][]{Brandt:2000,Giustini:2008}.

\section{Radiation-Driven Outflows from Large-Scale Inflows}

In the previous section, we showed that the powerful radiation from
an AGN can drive a strong wind from an accretion disk, the place
where the radiation is produced. In turn, this wind has a significant impact
on the radiation. However,
the same radiation can also
change the dynamics of the material at large distances from
the  radiation source. 

\begin{figure}[!ht]
\begin{center}
\includegraphics[width=7cm, angle=270]{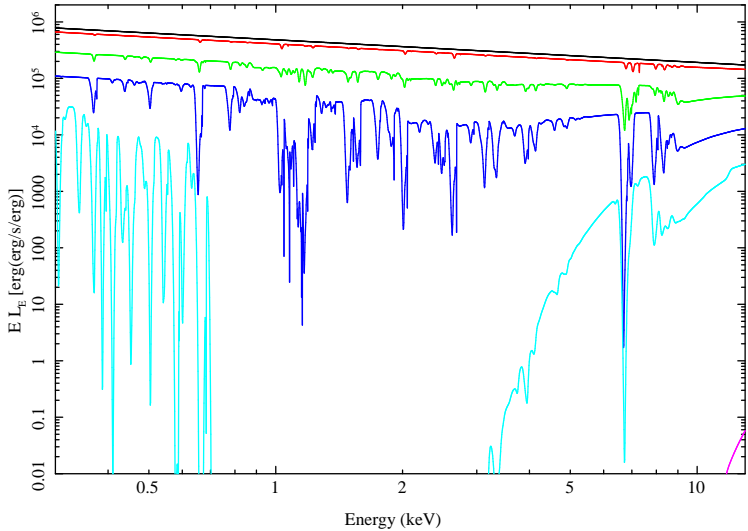}
\end{center}
\caption{
{\small The 0.3-13 keV transmitted X-ray spectra ($E\,L_{E}$) 
from five different lines-of-sight, with different inclinations, 
through the hydrodynamic simulation of an AGN outflow presented in \citet{Proga:2004}. 
The spectra were calculated with XSCORT ~(v5.18) using snapshot 800 of the 
\citeauthor{Proga:2004} simulation to provide self-consistent physical outflow properties. 
The input ionizing power-law spectrum is shown in black. The spectra 
correspond to inclinations of $\theta$ = 50, 57, 62, 65, \& 67$^\circ$,
(top to bottom). These l.o.s were 
chosen to highlight the range of spectral shapes that result from the 
different physical properties throughout the simulated wind. 
The figure is from \citet{Schurch:2009}.
}}
\end{figure}

We have begun studying gas dynamics in AGNs on subparsec and parsec scales 
\citep{Proga:2007b, Proga:2008, Kurosawa:2008, 
Kurosawa:2009a, Kurosawa:2009b, Kurosawa:2009c}. 
In the following, we briefly
summarize the main results of this work.

In \citet{Proga:2007b}, we calculated a series of models for non-rotating flows 
that are under the influence of super massive BH gravity and
radiation from an accretion disk surrounding
the BH.  Generally, we used the numerical methods developed by 
\citet{Proga:2000}.
Our numerical approach allows for the self-consistent
determination of whether the flow is gravitationally captured by the BH
or driven away by thermal expansion or radiation pressure.

For a $10^8~\MSUN$ BH with an accretion luminosity of 0.6 of $L_{\rm Edd}$
(the same parameters as in the disk wind simulations),
we found that a non-rotating flow settles 
quickly into a steady state and has two
components
(1) an equatorial inflow and
(2) a bipolar inflow/outflow with the outflow leaving the system
along the pole.
The first component is a realization of Bondi-like accretion flow. The
second component is an example of a non-radial accretion flow becoming
an outflow once it is pushed close to the rotational axis of the disk
where thermal expansion and radiation pressure can accelerate the flow
outward. The main result of these simplified calculations is that the
existence of the above two flow components is robust yet their
properties are sensitive to the geometry, SED of the radiation field,
and the outer boundaries.  In particular, the outflow power and the
degree of collimation are higher for the model with radiation
dominated by UV/disk emission than for the model with radiation
dominated by X-ray/central engine emission.  This sensitivity is
related to the fact that thermal expansion drives a weaker and wider
outflow, compared to the radiation pressure.

\begin{figure}[!ht]
\begin{center}
\includegraphics[width=7.5cm, angle=0]{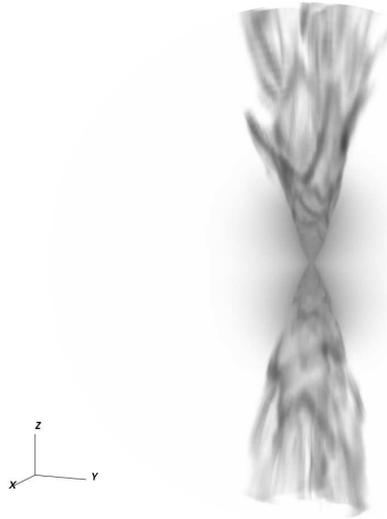}
\end{center}
\caption{
{\small 
 Three-dimensional hydrodynamical simulations of outflow
  formation via redirection of accreting gas by the strong radiation
  from an accretion disk around a super massive black hole with its
  mass $M_{\rm{BH}}=10^{8} \MSUN$.  The infalling gas is weakly
  rotating (sub-Keplerian), and the Eddington ratio of the system is 0.6.
  The volume rendering representation of the density distributions is
  shown. The outflow morphology is bi-conical, but the flow contains
  relatively cold and dense cloud-like structures which resembles
  those observed in the NRLs of Seyfert galaxies. 
The figure is from \cite{Kurosawa:2009b}.
}}
\end{figure}

Rotation of the inflowing gas changes the geometry of the flow because the
centrifugal force prevents gas from reaching the rotational axis (see
Fig.~1 in \citealt{Proga:2008}).  This, in turn, reduces the mass outflow
rate because less gas is pushed toward the polar region.  We also
found that rotation can lead to fragmentation and time variability of
the outflow. As the flow fragments, cold and dense clouds form
(Figs.~4 and 12 in \citealt{Proga:2008}).

Three-dimensional effects are also important.
In \cite{Kurosawa:2008}, we considered effects of radiation due to
a precessing accretion disk on a spherical cloud of gas around the disk.
On the other hand, in \cite{Kurosawa:2009a}, we recalculated some models
from papers \cite{Proga:2007b} and \cite{Proga:2008} in full 3-D.
Our 3-D simulations of a nonrotating gas show small yet noticeable
nonaxisymmetric small-scale features inside the outflow. However,
the outflow as a whole and the inflow do not seem to suffer from
any large-scale instability. In the rotating case, the nonaxisymmetric
features are very prominent, 
especially in the outflow which consists of many
cold dense clouds entrained in a smoother hot flow (e.g., see 
Figs.~4 and 5). 
The 3-D outflow is nonaxisymmetric due to the shear and thermal instabilities.
Effects of gas rotations are similar in 2-D and 3-D. In particular, 
gas rotation increases the outflow thermal
energy flux, but reduces the outflow mass and kinetic energy fluxes.
In addition,
rotation leads to time variability and fragmentation of the outflow
in the radial and latitudinal directions. The collimation of the outflow
is reduced in the models with gas rotation. The main difference between
the models with rotation in 3-D and 2-D is that
the time variability in the mass
and energy fluxes is reduced in the 3-D case because of
the outflow fragmentation in the azimuthal direction.

\begin{figure}[!ht]
\begin{center}
\includegraphics[width=7.5cm, angle=0]{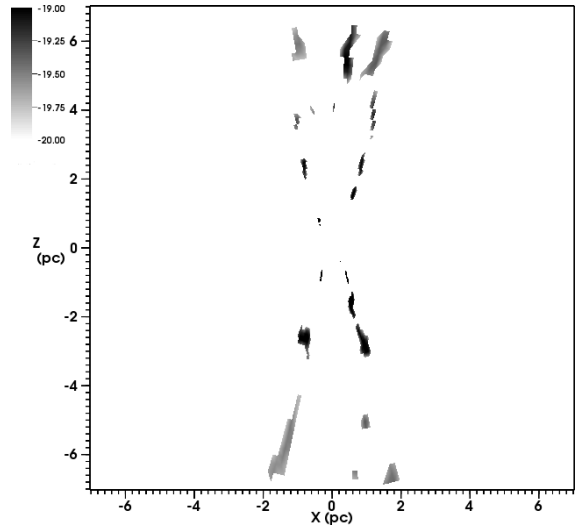}
\end{center}
\caption{
{\small 
Spatial distributions of the {}``cold clouds''
in the model shown in Fig. 4.
The grayscale image
shows the density map of the cold clouds in logarithmic scale (in cgs
unit) on the
$z$--$x$ plane. The cold clouds here are defined as the gas with
its density higher than
$\rho_{\mathrm{min}}=1.6\times10^{-20}\,\mathrm{g\, cm^{-3}}$ and
its temperature less than $T_{\mathrm{max}}=1.6\times10^{5}\,\mathrm{K}$.
The clouds are not spherically distributed, but located near the bi-conic
surface (which appears as an X-shaped pattern here) defined by the outflowing
gas. Note that the length scale are in units of pc. 
The figure is from \citet{Kurosawa:2009b}.
}}
\end{figure}

To be able to compare these new simulations with observations,
we are in a process of computing synthetic line profiles, broad band
spectra and maps. However, even without these diagnostics we can
check if the models are consistent with the data.
For example, we can compare the kinematics study of NGC~4151 \citet{das:2005}
with
the velocity of the cold clouds 
(cf.~Fig.~5) projected ($v_{\mathrm{proj}}$)
toward an observer at the inclination angle $i=45^{\circ}$
, which
is the inclination of NGC~4151.
\citet{das:2005}
used  the kinematics model of the outflows with a bi-conic radial
velocity law, and found a good fit to their observations
when the opening angle of the cone is $\sim 33^{\circ}$. Interestingly,
we find the opening angle of the outflows in our is also about
$30^{\circ}$ (cf.~Figs.~5).

Figure~6 shows $v_{\mathrm{proj}}$ of the
clouds plotted as a function of the projected vertical distance, which
is the distance along the $z$-axis in Fig.~5
projected onto the plane of the sky for an observer viewing the
system with $i=45^{\circ}$. The figure shows that the clouds are
accelerated up to $250\,\mathrm{km\, s^{-1}}$ until the projected
distance reaches $\sim4$~pc, but the velocity curve starts to flatten
beyond this point. Towards the outer edges (near the outer boundaries),
the curve begins to show a sign of deceleration, but not so clearly.
We note that the hot outflowing
gas, on the other hand, does show deceleration at the larger radii
in our models \citep*[cf.~Fig.~9 in][]{Kurosawa:2009a}.
Although the physical size of the long slit observation of NGC~4151
by \citet{das:2005} is in much lager scale ($\sim50$ times larger)
than that of our model, their radial velocities as a function of the
position along the slit (see their Figs.~5 and 6) show a similar
pattern as in our model (Fig\@.~6). The
range of $v_{\mathrm{proj}}$ in our model 
is about $-250$ to $300\,\mathrm{km\,\ s^{-1}}$
while the range of the observed radial velocities in \citet{das:2005}
is about $-800$ to $800\,\mathrm{km\, s^{-1}}$, which is comparable
to ours. 

\begin{figure}[!ht]
\begin{center}
\includegraphics[width=8cm, angle=0]{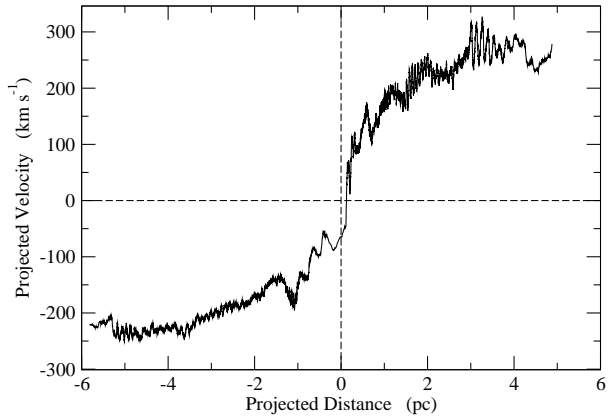}
\end{center}
\caption{
{\small 
The velocities of the cold cloud elements
(as in Fig.~5) projected toward an observer located
at the inclination angle $i=45^{\circ}$ are shown as a function of
the projected vertical distance (the distance along the $z$-axis
in Fig.~5, but projected on to the plane of the
sky for the observer viewing the system with $i=45^{\circ}$). The
negative projected distance indicates the clouds are found in the
lower half of the projection plane. 
The figure is from \citet{Kurosawa:2009b}.
}}
\end{figure}

\section{Concluding Remarks}

The models presented here, which numerically simulate the outflows driven by
radiation from AGN, are in many respects consistent with
observations. Clearly more work 
is needed to test the models and improve them. However,
the first few steps toward the development a self-consistent physical model
of the AGN outflows has been taken. 

Applications of the model are not limited to AGN because other astrophysical
objects -- such as X-ray binaries, in particular micro-quasars -- 
have a similar geometry and can be understood within a similar physical
framework. In addition, having a physical model of the AGN outflows
we can apply it to the so-called AGN feedback problem.

Some results from the outflow models have been already incorporated 
to galaxy evolution calculations. In particular, \citet{Ciotti:2009}, 
improved and extended the accretion and feedback physics 
explored in their previous papers \citep[e.g.,][]{Ciotti:1997, 
Ciotti:2001,Ciotti:2007}.
By using a high-resolution one-dimensional hydrodynamical code, 
\citet{Ciotti:2009} studied,
the evolution of an isolated elliptical galaxy,
where the cooling and heating functions include photoionization 
and Compton effects, and restricting to models which include 
only radiative or only mechanical feedback 
\citep[the latter, in the form of disk winds properties of which
were adopted from][]{Proga:2000, Proga:2004}. 

These recent calculations confirmed that for Eddington ratios above
0.01, both accretion and radiative outputs are forced by feedback effects to be 
in burst mode, so that strong intermittencies are expected at early times, 
while at low redshift the explored models are characterized by smooth, 
very sub-Eddington mass accretion rates punctuated by rare outbursts. 
However, the explored models always fail some observational tests. 
For the high mechanical efficiency of $10^{-2.3}$  as 
adopted by some investigators, it was found that most of the gas 
is ejected from the galaxy, the resulting X-ray luminosity is less than 
a value typically observed and little super massive black hole 
growth occurs. However, models with low enough mechanical efficiency 
to accommodate satisfactory black hole growth tend to 
allow too strong cooling flows and leave galaxies at z = 0 
with E+A spectra more frequently than is observed. 
Surprisingly, it was also found that both types of feedback 
are required. Radiative heating over the inner few kilo parsecs is needed to 
prevent calamitous cooling flows, and mechanical feedback from 
active galactic nucleus winds, which affects 
primarily the inner few hundred parsecs, 
is needed to moderate the luminosity and growth of the central black hole. 
Models with combined feedback pass more observational tests
(Ciotti et al., in preparation). Thus it emerges from these simulations that
to solve the AGN feedback problem/explain all aspects of the galaxy evolution, 
more than one form of feedback is needed.

\acknowledgements %%% Text of acknowledgements runs on after this command.
This work was supported by NASA through grant HST-AR-11276 from the
Space Telescope Science Institute, which is operated by the Association
of Universities for Research in Astronomy, Inc., under NASA contract
NAS5-26555. DP acknowledges also support from NSF (grant AST-0807491).

%%% THE BIBLIOGRAPHY
%%%
%%% CONSULT SECTION 3 OF "INSTRUCTIONS FOR AUTHORS" FOR HOW TO USE NATBIB.
%%% AUTHORS ARE ENCOURAGED TO USE EITHER THE "THEBIBLIOGRAPY" ENVIRONMENT
%%% BY UNCOMMENTING (DELETING THE "%" SYMBOL) THE COMMANDS BELOW, OR BY
%%% USING THE BIBTEX ENVIRONMENT. TO FIND OUT WHICH IS APPLICABLE TO YOUR
%%% CONTRIBUTION, CONSULT THE VOLUME EDITORS FOR YOUR PROCEEDINGS.
%%%


\begin{thebibliography}
\expandafter\ifx\csname natexlab\endcsname\relax\def\natexlab#1{#1}\fi

\bibitem[{Brandt {et~al.}(2000)}]{Brandt:2000}
Brandt, W.~N., Laor, A., \& Wills, B.~J. 2000, ApJ, 528, 637

\bibitem[{Castor {et~al.}(1975)Castor, Abbott, \& Klein}]{Castor:1975}
Castor, J.~I., Abbott, D.~C., \& Klein, R.~I. 1975, \apj, 195, 157

\bibitem[{Ciotti \& Ostriker(1997)}]{Ciotti:1997}
Ciotti, L. \& Ostriker, J.~P. 1997, \apjl, 487, L105

\bibitem[{Ciotti \& Ostriker(2001)}]{Ciotti:2001}
---. 2001, \apj, 551, 131

\bibitem[{Ciotti \& Ostriker(2007)}]{Ciotti:2007}
---. 2007, \apj, 665, 1038

\bibitem[{Ciotti {et~al.}(2009)Ciotti, Ostriker, \& Proga}]{Ciotti:2009}
Ciotti, L., Ostriker, J.~P., \& Proga, D. 2009, \apj, 699, 89


\bibitem[{Crenshaw {et~al.}(2002)Crenshaw, Kraemer, \& George}]{Crenshaw:2002}
 Crenshaw, D.M., Kraemer, S.B. \& George I.M.
  Mass Outflow in Active Galactic Nuclei: New Perspectives,
  ASP Conference Proceedings, Vol. 255. Edited by D. M. Crenshaw,
   S. B. Kraemer, and I. M. George, San Francisco: Astronomical Society
  of the Pacific, 2002

\bibitem[{Das {et~al.}(2005)Das, {Crenshaw}, {Hutchings}, {Del}, {Kraemer},
  {Gull}, {Kaiser}, {Nelson}, \& {Weistrop}}]{das:2005}
Das, V. et al. 2005, \aj, 130, 945

\bibitem[{Giustini {et~al.}(2008)}]{Giustini:2008}
Giustini, M., Cappi, M., \& Vignali, C. 2008,
A\&A, 491, 425

\bibitem[{K$\ddot{\rm o}$nigl}(2006)]{Konigl:2006}
K$\ddot{\rm o}$nigl A. 2006, MmSAI,77, 598

\bibitem[{Kurosawa \& Proga(2008)}]{Kurosawa:2008}
Kurosawa, R. \& Proga, D. 2008, \apj, 674, 97

\bibitem[{Kurosawa \& Proga(2009{\natexlab{a}})}]{Kurosawa:2009a}
---. 2009{\natexlab{a}}, \apj, 693, 1929

\bibitem[{Kurosawa \& Proga(2009{\natexlab{b}})}]{Kurosawa:2009b}
---. 2009{\natexlab{b}}, \mnras, 397, 1791

\bibitem[{Kurosawa {et~al.}(2009)}]{Kurosawa:2009c}
Kurosawa, R., Proga, D. \& Nagamine K. 2009, \apj, 707, 823

\bibitem[{{Proga}(1999)}]{Proga:1999b}
{Proga}, D. 1999, \mnras, 304, 938

\bibitem[{Proga}(2005)]{Proga:2005}
 Proga, D. 2005,  
 in ASP Conf. Ser. 330,  The Astrophysics of Cataclysmic Variables and
 Related Objects, ed. J.-M. Hameury \& J.-P. Lasota (San Francisco: ASP),
 103

\bibitem[{Proga}(2007a)]{Proga:2007a}
Proga, D. 2007a, in The Central Engine of Active Galactic Nuclei, 
ed. L. C. Ho, \& J.-M. Wang, ASP Conf. Ser., 373, 267

\bibitem[{Proga(2007b)}]{Proga:2007b}
Proga, D. 2007b, \apj, 661, 693

\bibitem[{{Proga} \& {Kallman}(2004)}]{Proga:2004}
{Proga}, D. \& {Kallman}, T.~R. 2004, \apj, 616, 688

\bibitem[{Proga {et~al.}(2008)Proga, {Ostriker}, \& {Kurosawa}}]{Proga:2008}
Proga, D., {Ostriker}, J.~P., \& {Kurosawa}, R. 2008, \apj, 676, 101

\bibitem[{Proga {et~al.}(1998)Proga, Stone, \& Drew}]{Proga:1998}
Proga, D., Stone, J.~M., \& Drew, J.~E. 1998, \mnras, 295, 595

\bibitem[{Proga {et~al.}(1999)Proga, Stone, \& Drew}]{Proga:1999a}
Proga, D., Stone, J.~M., \& Drew, J.~E. 1999, \mnras, 310, 476

\bibitem[{Proga {et~al.}(2000)Proga, Stone, \& Kallman}]{Proga:2000}
Proga, D., Stone, J.~M., \& Kallman, T.~R. 2000, \apj, 543, 686

\bibitem[{Puls {et~al.}(2008)}]{Puls:2008}
Puls, J., Vink, J. S., \& Najarro, F. 2008, 
A\&AR, 16, 209

\bibitem[{Richards \& Hall}(2004)]{Richards:2004}
Richards, G.T. \& Hall, P.B 2004, AGN Physics with the Sloan Digital Sky 
Survey, ASP Conference Series, Volume 311,
Edited by Gordon T. Richards and  Patrick B. Hall, 
San Francisco: Astronomical Society of the Pacific, 2004

\bibitem[{Schurch {et~al.}(2009)}]{Schurch:2009}
Schurch, N.~J., Done, C., \& Proga, D, 2009, \apj, 694, 1

\end{thebibliography}
\end{document}